\documentstyle[psfig]{l-aa}
\begin{document}

\thesaurus{01(08.02.5, 08.09.2 (StHA 190), 09.10.1)}

\title{{\it Letter to the Editor} \\ 
	Discovery of a bipolar and highly variable mass outflow from the symbiotic 
        binary StH$\alpha$~190\thanks{Based in part on observations secured with
        SARG at Telescopio Nazionale Galileo (TNG), La Palma, Canary Islands}}

\author{U.~Munari\inst{1,2} 
   \and T.~Tomov\inst{2,3} 
   \and B.F.~Yudin\inst{4}
   \and P.M.~Marrese\inst{1,2}
   \and T.~Zwitter\inst{5}
   \and R.G.~Gratton\inst{6}
   \and G.~Bonanno\inst{7}
   \and P.~Bruno\inst{7}
   \and A.~Cal\'{i}\inst{7}
   \and R.U.~Claudi\inst{6} 
   \and R.~Cosentino\inst{7,8}
   \and S.~Desidera\inst{6}
   \and G.~Farisato\inst{6}
   \and G.~Martorana\inst{1}
   \and G.~Marino\inst{8}
   \and M.~Rebeschini\inst{1}
   \and S.~Scuderi\inst{7}
   \and M.C.~Timpanaro\inst{7}}

\institute{Osservatorio Astronomico di Padova, Sede di Asiago, I-36032 Asiago
           (VI), Italy 
\and       Center of Studies and Activities for Space (CISAS ``G.Colombo"), Uniersity of Padova,
	   Italy 
\and       Center for Astronomy, Nicolaus Copernicus University, ul. Gagarina 11, PL-87100 Torun, Poland
\and       Sternberg Astronomical Institute, Universitetskii pr. 13, Moscow 119899, Russia
\and       Department of Physics, Univ. of Ljubljana, Jadranska 19, 1000 Ljubljana, Slovenia
\and       Osservatorio Astronomico di Padova, Vicolo dell'Osservatorio 5,
           I-35122 Padova, Italy
\and       Osservatorio Astrofisico di Catania, Via S.Sofia 78,  I-95123 Catania, Italy 
\and       Centro Galileo Galilei - CNAA, Calle Alvarez de Abreu 70,
           38700 Santa Cruz de La Palma (TF), Spain
}

\date{Received date..............; accepted date................}

\maketitle

\markboth{U.~Munari et al.: Bipolar mass outflow from the symbiotic star 
StH$\alpha$ 190}{U.~Munari et al.: Bipolar mass outflow from the symbiotic 
star StH$\alpha$ 190}

\begin{abstract}
A highly and rapidly variable bipolar mass outflow from StH$\alpha$~190 has
been discovered, the first time in a yellow symbiotic star. Permitted
emission lines are flanked by symmetrical jet features and multi-component
P-Cyg profiles, with velocities up to 300 km~sec$^{-1}$. Given the high
orbital inclination of the binary, if the jets leave the system nearly
perpendicular to the orbital plane, the de-projected velocity equals or
exceeds the escape velocity (1000 km~sec$^{-1}$). StH$\alpha$~190 looks
quite peculiar in many other respects: the hot component is an O-type
sub-dwarf without an accretion disk or a veiling nebular continuum and the
cool component is a G7~III star rotating at a spectacular 105 km~sec$^{-1}$,
unseen by a large margin in field G giants.
\end{abstract}
\maketitle

\section{Introduction}

StH$\alpha$~190 has been independently discovered on objective prism plates
by Kinman (1983, private communication to Whitelock et al. 1995, hereafter
W95) and Stephenson (1986). Its symbiotic nature was noted
during the spectroscopic survey of StH$\alpha$ objects by Downes \& Keyes
(1988). The 3300-9100 \AA\ absolutely fluxed spectrum of StH$\alpha$~190 
included in the spectrophotometric atlas of 137 symbiotic stars by Munari 
\& Zwitter (2001, hereafter MZ01) shows a well developed G-type continuum with 
minimal veiling by the nebular or hot companion continua and a pronounced 
emission line spectrum of moderate excitation (HeII missing) with strong [OIII]
and [NeIII] forbidden lines. Minimal - if any - changes arise in the
comparison with older available spectroscopy. The IUE spectrum of
StH$\alpha$~190 by Schmidt \& Nussbaumer (1993, hereafter SN93) confirms the
moderate excitation conditions (NV and HeII missing) and shows the
photospheric continuum of an O sub-dwarf without contribution from nebular
regions or an accretion disk.

Data from the Munari et al. (2001, hereafter MHZ) {\sl UBV(RI)$_C$}
photometric survey of symbiotic stars confirms a rather limited variability
of StH$\alpha$~190 and absence of outbursts since its discovery, reporting
$V$=10.50, $(B-V)=+$0.84, $(U-B)=-$0.23, $(V-R)_C=+$0.50 and
$(V-I)_C=+$0.47 for mid 2000. W95 infrared photometry of StH$\alpha$~190 
over 16 nights from Oct. 1983 to Jul. 1987 gives $K=7.81$, $J-H=+0.57$, 
$H-K=+0.36$ and $K-L=+0.95$ as mean values. In $J$ and $H$ bands W95 did 
not find evidence of variability while a modest $\bigtriangleup K = 0.16$ 
is attributed to changes in the heating of the circumstellar dust by the hot source.

In this {\sl Letter} we report about the discovery of highly variable
bipolar mass outflow and blob ejection from StH$\alpha$~190, the first time
in a {\sl yellow} symbiotic star. Yellow SS harbour warm giants (F,G or
early K type), which have much smaller dimensions and lower mass loss rates
compared to the M giants of classical symbiotics. Jets and bipolar outflows
have been so far discovered in only five other symbiotic stars (among the
$\sim$225 known), all of them containing M giants or Miras and showing
outburst activity: R~Aqr (Burgarella \& Paresce 1992), CH~Cyg (Taylor et al.
1986), MWC~560 (Tomov et al. 1990), RS~Oph (Taylor et al. 1989) and
Hen~3-1341 (Tomov et al. 2000).

\begin{figure}[!t]
\centerline{\psfig{file=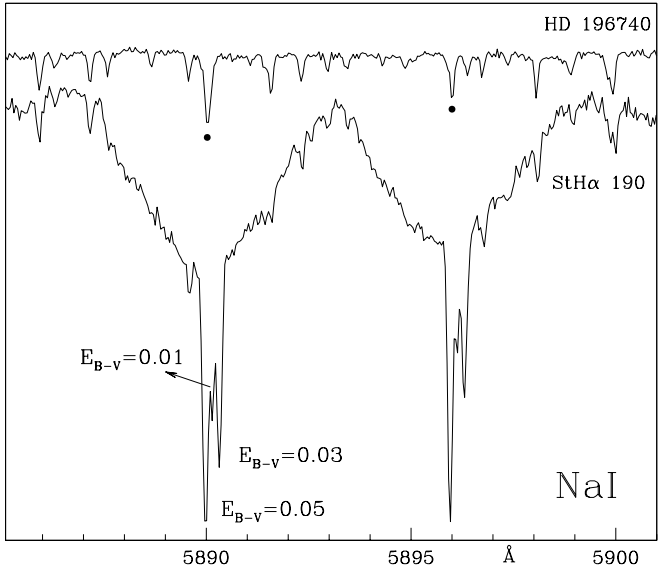,width=8.8truecm}}
\caption[]{The region around the NaI~doublet, showing the rotationally
           broadened stellar component and three sharper interstellar lines
           (SARG spectrum for Oct 9, 2000).  The upper spectrum shows the
           telluric absorptions in 28~Vul, a B6~IV fast rotating star
           (V$_{\rm rot}$sin$i$=320 km sec$^{-1}$) observed on the same night as
           StH$\alpha$~190. 28~Vul suffers from a $E_{B-V} \sim 0.01$
           responsible for the interstellar NaI lines marked by dots.
           Comparing 28~Vul and StH$\alpha$~190 spectra reveals how nearly
           all the sharp and weak lines in the latter spectrum are
           telluric.}
\end{figure}
\begin{table}[!t]
\centerline{\psfig{file=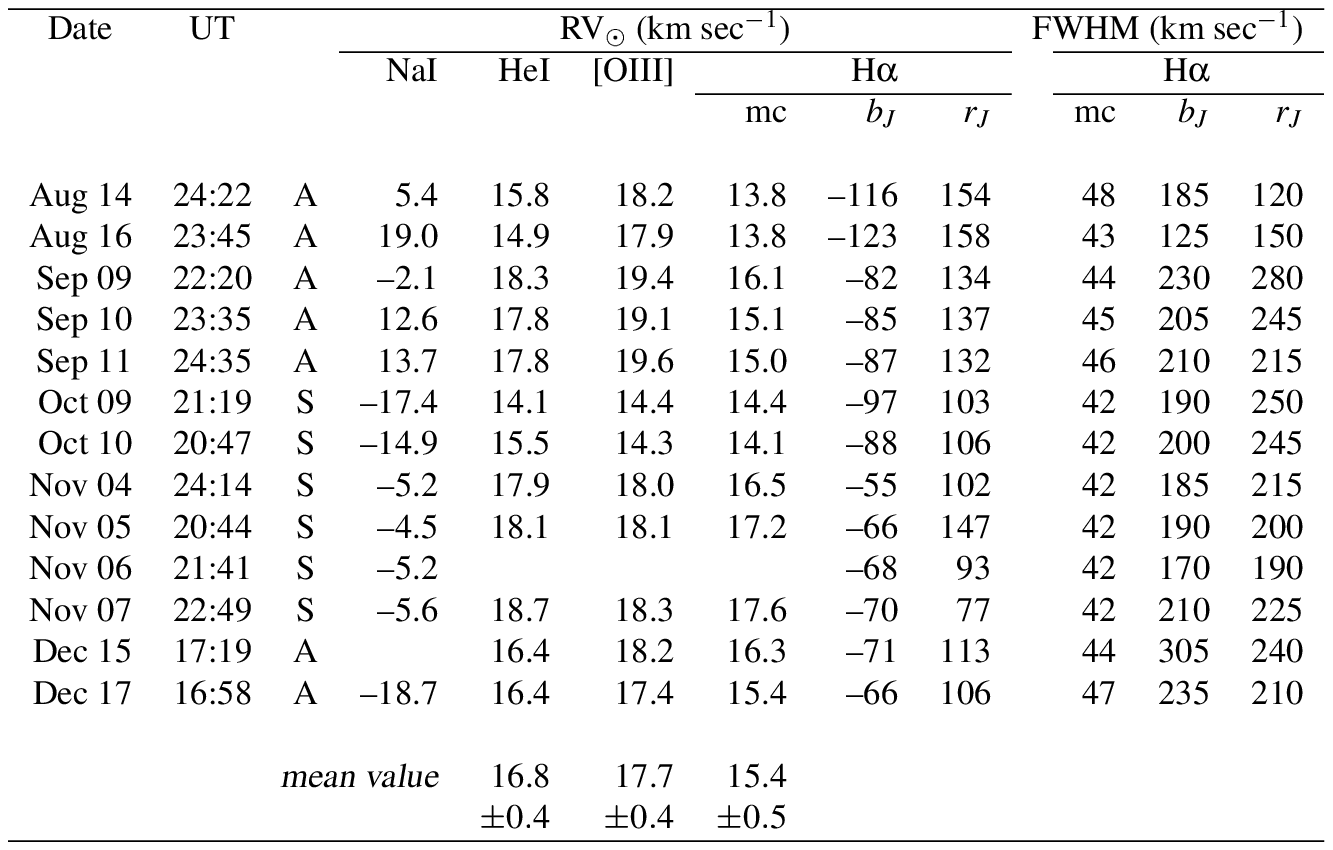,width=8.8truecm}}
\caption[]{Journal of spectroscopic observations and measured radial
           velocities. $A$= Asiago Echelle+CCD spectrograph operating at
           $R$=18,500 resolving power, $S$= the SARG Echelle+CCD spectrograph at
           $R$=57,000. The heliocentric radial velocities of HeI~5876 and
           [OIII]~4959 \AA\ pertain to the main emission component, those of
           NaI to the stellar 5890-5896 doublet de-blended from the
           interstellar components. Data on H$\alpha$ includes heliocentric
           radial velocities of the main component ({\sl mc}), the blue
           shifted jet ($b_J$) and the red shifted jet ($r_J$), as well as
           the component widths corrected for the instrumental PSF.}
\end{table}

\section{Observations}

$R=\frac{\lambda}{\bigtriangleup \lambda}=18,500$ spectra have been obtained with 
the Echelle spectrograph mounted at the Cassegrain focus of the 1.82\,m
telescope which is operated by the Padova and Asiago Astronomical Observatories
on top of Mt.~Ekar, (Asiago, Italy). The detector has been a Thomson THX31156
CCD with 1024$\times$1024 pixels, 19$\mu$m each, and the slit width has been
set to 1.8 arcsec. The spectra cover the 4500-9000 \AA\ range.

$R=\frac{\lambda}{\bigtriangleup \lambda}=57,000$ spectra have been obtained
during the commissioning phase of the SARG white-pupil Echelle spectrograph
for the Italian 3.5 m Telescopio Nazionale Galileo (La Palma, Canary
Islands). The detector is a mosaic of two $4k\times 2k$\ thinned, back
illuminated EEV CCDs. A 0.80~arcsec slit width has been used together with
the {\sc Yellow} cross-disperser grism which provides a nearly complete
spectral coverage from 4650 to 7900~\AA.
A journal of the observations is given in Table~1.

Infrared photometry has been obtained at the 1.25 m telescope of the Crimean
Astrophysical Observatory, giving $J$=8.74, $H$=8.25, $K$=8.08, $L$=7.48 on
Nov. 20 and $J$=8.74, $H$=8.22, $K$=8.04, $L$=7.52 on Dec. 8, 2000 (with errors 
$\pm 0.02$ in JHK, $\pm 0.04$ in L).

\section{System properties}

\subsection{Classification, distance, interstellar lines and reddening}

The 3300-9100 \AA\ spectra of StH$\alpha$~190 and MKK standards in MZ01
allows us to classify the cool giant as G7~III. Comparing MHZ photometry
with the intrinsic colours from Fitzgerald (1970) and absolute magnitudes
from Schmidt-Kaler (1982), the reddening turns out to be $E_{B-V}$=0.10 and
the distance $d=575$ pc. SN93 estimated an identical $E_{B-V}$=0.10 from the
2175\AA\ interstellar hump in the IUE spectra of StH$\alpha$~190.

Multiple interstellar components are superposed on the rotationally
broadened stellar NaI doublet (cf. Figure~1). Their RV$_\odot$ are $-18.1\
(\pm 0.5)$, $-10.0\ (\pm 0.5)$ and $-0.9\ (\pm 0.4)$ km~sec$^{-1}$, with
0.159, 0.051 and 0.112 ($\pm$0.003) \AA\ as equivalent widths for the 5889
component, respectively.  They are unresolved on the Asiago spectra and the
blend has RV$_\odot = -10.8\ (\pm 0.5)$ km~sec$^{-1}$. StH$\alpha$~190 is at
$b=-36^\circ$ so our line-of-sight exits the galactic dust layer (assumed to
reach $\bigtriangleup z \sim 100$ pc over the galactic plane) at the
projected distance of $\sim$170 pc, where the effect of the galactic
rotation on the radial velocity does not exceed 2 km~sec$^{-1}$. None of the
three interstellar lines shares the velocity of the StH$\alpha$~190
circumstellar material (see bottom line in Tab.~1), so they have to
originate in distinct clouds with RV dispersion similar to that of extreme
Pop~I objects (12.5 km~sec$^{-1}$, Binney \& Merrifield 1998). If we use the
NaI vs. $E_{B-V}$ calibration of Munari \& Zwitter (1997), the equivalent
widths of the three interstellar NaI components correspond to
$E_{B-V}=0.049$, 0.016 and 0.035, respectively, giving a total
$E_{B-V}=0.10$, a value identical to what above derived by independent
methods.

\begin{figure*}[!ht]
\centerline{\psfig{file=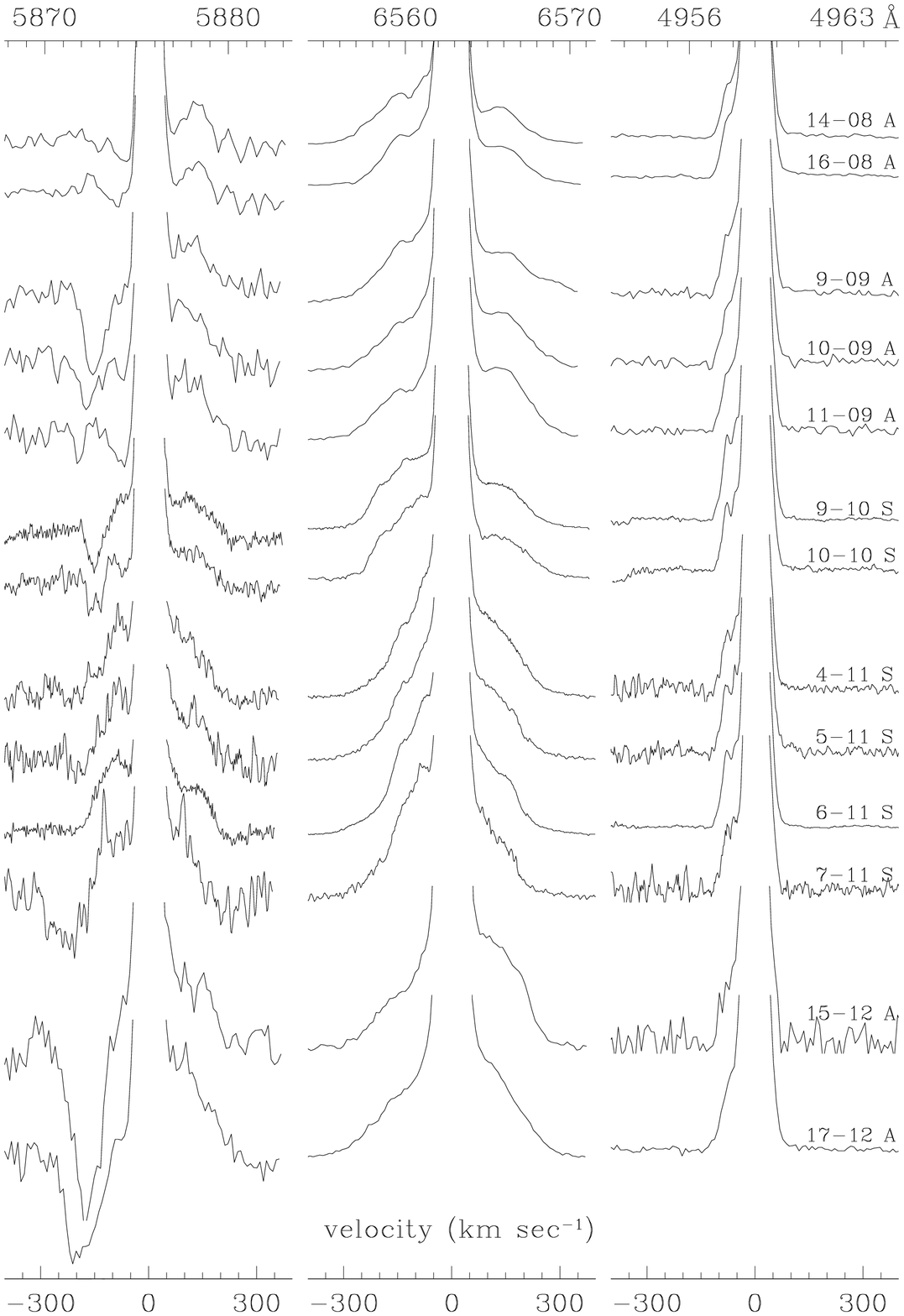,width=15.0truecm}}
\caption[]{Evolution of the HeI, H$\alpha$ and [OIII] emission line profiles
           over a 4 month period (see dates on the right and Tab~1); $S$= SARG,
           $A$=Asiago spectra.  The higher resolution of SARG spectra is the
           reason for the {\em apparent} variability of the [OIII] line,
           which is actually pretty constant. The vertical scale is constant for
           each line (expansion factors 1:7.2:7.2), the continuum is normalised
           and the line profiles are truncated.}
\end{figure*}

\subsection{Radial Velocities and Orbital Motion}

Heliocentric velocities of the main emission lines and the stellar NaI
absorption which traces the G7~III star are listed in Table~1. The fairly
constant RV$_\odot$ of emission lines indicates that their formation regions
do not follow the orbital motion and are circumstellar in origin.

The stellar NaI doublet clearly shows orbital motion velocity shifts, even
if the observations did not cover a full orbital cycle.  Lower limit to the
velocity amplitude ($\bigtriangleup RV_\odot \sim$30 km sec$^{-1}$) is
remarkable for a symbiotic star and it is the largest recorded so far (see
Tab.~4 of Belczy\'{n}ski et al. 2000). This suggests a large orbital
inclination, an unusually massive hot companion and a short orbital period.

\subsection{Photometric variability}

StH$\alpha$~190 has been detected by Tycho at the limit of its sensitivity
range during 78 passages distributed over 17 dates (from Dec 27, 1989 to Dec
14, 1992). Automatic analysis of Tycho data summarized in the Hipparcos
Catalogue did not detected variability of StH$\alpha$~190 over the large
noise in the $B_T$ and $V_T$ data. Our recent IR photometry reported in
Sect. 2 is in excellent agreement with the older W95 data.

However, if a detailed search for variability and periodicities is performed
on Tycho and W95 data as well as on the radial velocities of Table~1, an
average periodicity of 171~$\pm$5 days (and its 115~$\pm$7 yearly alias) is
found. If this corresponds to the orbital period it would be 
the shortest known among all symbiotic stars, followed by those of two other 
yellow symbiotics TX CVn (199 days) and BD-21.3873 (282 days) and the
recurrent symbiotic nova T~CrB (228 days; cf. Belczy\'{n}ski et al. 2000
and references therein).

\subsection{Rotational velocity}

The width of the stellar NaI absorption lines in Figure~1 corresponds to
$V_{\rm rot}$ sin$\,i \sim$105 km sec$^{-1}$, which translates into a 5 days
rotation period for the G7~III star, much less than the possible 171 day
orbital period.  Such a rotational velocity is very high: from the catalogue
of rotational velocities of Bernacca \& Perinotto (1973) the mean value for
the 288 giants between G2 K2 is $V_{\rm rot}$=9.7 km~sec$^{-1}$, with 92\%
of them having $V_{\rm rot}\leq$10 km~sec$^{-1}$. The high $V_{\rm rot}$
sin$\,i$ further strengthens the idea of a high orbital inclination for
StH$\alpha$~190.

\section{The bipolar jets and blobby mass outflow}

Figure~2 presents the temporal evolution of HeI 5876 \AA\ emission line
profile as a template for other helium lines, H$\alpha$ for hydrogen, and
[OIII] for the nebular lines. Over the observational period no other
substantial change affected the spectrum of StH$\alpha$~190.

The H$\alpha$ profile is dominated by a central component that has remained
remarkably constant over the last four mounts (see Table~1). Two weaker and
symmetrically placed components are flanking the central component. They
show large day-to-day variability in both $RV_\odot$ and width (cf.
Table~1). We identify them as spectral signatures of jet-like discrete
ejection events. Weak P-Cyg absorptions interfere with the blue jet
component, reducing its width and its velocity shift vs. the main H$\alpha$
component. Orbital inclination of StH$\alpha$~190 is probably high, so the
de-projected velocity of the jet components must be much larger than the
observed velocity shifts ($\sim$150 km~sec$^{-1}$) and well in excess of the
escape velocity from the O sub-dwarf companion to the G7~III ($\sim$1000
km~sec$^{-1}$). The mass of the gas originating the jets is 
$10^{-11} \ \ M_\odot$, while the circumstellar ionized region
has $M = 1\times 10^{-6} \ M_\odot$ and $R = 4.5\times 10^{14}$ cm = 30 AU 
(assuming a simple spherically symmetric geometry). The mass
loss rate necessary to sustain the jets is {\sl \.M} $\sim 5\ 10^{-8}\
(V_{\rm jet}/1000\ {\rm km~sec^{-1}})\ \ M_\odot\ yr^{-1}$.

The jet components are visible in the profiles of HeI lines too (see
HeI~5876 \AA\ in Figure~2). The velocity and profile of the red component
corresponds closely to that seen in the H$\alpha$ profile. The most
outstanding feature of the HeI profile is however the multi-component and
highly variable P-Cyg component, with terminal velocity even in excess of
300 km~sec$^{-1}$. The P-Cyg absorption can be so strong as to completely
overwhelm the jet's blue component. The P-Cyg profiles evolve on a few days
or hours time scale. The strong P-Cyg component on Sept.~9 (see Fig.\ 2),
for example, accelerated outward by 20 km~sec$^{-1}$ day$^{-1}$ and
dissolved in the next two days, while being replaced by a growing and
accelerating new P-Cyg component. The typical mass involved in the
absorption profiles is $M = 10^{-11}\, -\, 10^{-10} \ M_\odot$.

The relative appearance of Hydrogen and HeI lines in  StH$\alpha$~190
closely resembles Hen~3-1341 and its well developed jets and mass outflow
(Tomov et al. 2000). We are continuing with the photometric and
spectroscopic monitoring and a detailed modeling of StH$\alpha$~190 will be
presented elsewhere.

\end{document}